\documentstyle[aps,prl,preprint]{revtex}
\begin{document}
\draft
\title{Nucleosynthesis and the Mass of the $\tau$ Neutrino---Erratum}
\author{Steen Hannestad and Jes Madsen}
\address{Institute of Physics and Astronomy,
University of Aarhus, 
DK-8000 \AA rhus C, Denmark}
\date{27 June 1996}
\maketitle

In a recent {\it Letter\/}\cite{han96} we presented the first numerical
treatment of the full set of Boltzmann equations for the evolution of an
MeV Majorana $\tau$ neutrino in the early Universe and concluded 
that mass limits
from Big Bang nucleosynthesis were significantly weakened compared to
previous investigations.

An error in our numerical code unfortunately
invalidates the results. As noticed by several people\cite{thanks}
one should expect a strengthening rather than a weakening
of earlier mass limits based on the integrated Boltzmann
equation\cite{kolb,dolrot} because our inclusion of 
the full neutrino spectra should reduce the annihilation of
high-momentum neutrinos, thereby permitting more $\tau$ neutrinos to
survive. This is indeed the case as shown in the revised Fig.\ 1\cite{kawa}. 
Our neutrino distribution functions now differ less from kinetic
equilibrium than found in\cite{han96}. A typical case for $\nu_\tau$
is shown in Fig.\ 2.
Deviations by a factor of 2 appear frequently when compared to a
kinetic equilibrium distribution,
$f_\nu =\left[\exp\left( (E-\mu)/T\right) +1\right] ^{-1}$,
with the same number density and $T=T_\gamma$, as often assumed, whereas the
distribution is close to that of a kinetic distribution with $T$ and
$\mu$ determined by the number- and energy densities.

The resulting $\nu_e$ and $\nu_\mu$ distributions are
still significantly heated relative to the standard case with 
a massless $\nu_\tau$. 
For eV-mass $\nu_\mu$ or $\nu_e$ this changes the present day
contribution to the cosmic density to $\Omega_\nu h^2=\alpha m_\nu
/93.03$eV  with $\alpha =1$ for a massless $\nu_\tau$, and $\alpha
=$1.10(1.13), 1.09(1.14), 1.03(1.05), 1.01(1.02) for $\nu_e$($\nu_\mu$)
for $m_{\nu_\tau}=$5, 10, 15, 20 MeV ($h$ is the Hubble-parameter in
units of 100km s$^{-1}$ Mpc$^{-1}$). Later decay of $\nu_\tau$ can
further increase the value of $\alpha$.

Fig.\ 3 illustrates the consequences for
Big Bang nucleosynthesis in terms of the equivalent number of massless
neutrinos, $N_{\text{eq}}$, needed to give a similar production of $^4$He at a
baryon-to-photon ratio $\eta =3\times 10^{-10}$. The revised results
with and without inclusion of the change in the $\nu_e$ number density
are in fine agreement with recent results
based on the integrated Boltzmann equation by Fields, Kainulainen, and
Olive\cite{fields}. Including also the actual shape of the $\nu_e$
distribution to some extent compensates for the effect on the energy density, a
result also found by Dolgov, Pastor, and Valle\cite{dolgov,kimmo}, though we
disagree by several equivalent neutrino species with the total $N_{\text{eq}}$
found in \cite{dolrot}, and to some extent also with the differential
changes in $N_{\text{eq}}$ quoted in \cite{dolgov}.

In conclusion our revised results show that no MeV $\tau$-neutrino
with mass below the experimental limit of 24 MeV is permitted unless 
more than 4 equivalent massless neutrino flavors become allowed by
future observations of the primordial element abundances.

\begin{figure}
\caption{The relic number density of massive $\tau$ neutrinos 
normalized to a massless species times mass, $r m_{\nu}$.
The solid curve is calculated using the full
Boltzmann equation and all interactions. Dotted and dashed curves
are adopted from \protect\cite{kolb} and \protect\cite{dolrot}.}
\label{fig1}
\end{figure}

\begin{figure}
\caption{The distribution of $\tau$ neutrinos
of mass 15 MeV at a photon temperature of 0.92 MeV.
The solid curve includes all possible interactions whereas the
dashed-dotted curve only includes annihilations. The dotted curve is a
kinetic equilibrium distribution with the same number density 
as that of the solid curve and $T=T_\gamma$,
whereas the dashed curve has the same number- and energy densities as
the solid, with $T$ and $\mu$ adjusted accordingly.}
\label{fig2}
\end{figure}

\begin{figure}
\caption{Equivalent number of massless neutrinos, shown as
$\Delta N_{\text{eq}} = N_{\text{eq}}-3$. The dashed
curve is without heating of $\nu_e$, the dotted with
only the electron neutrino number density changed, and the
solid curve with the full distribution of $\nu_e$.}
\label{fig3}
\end{figure}

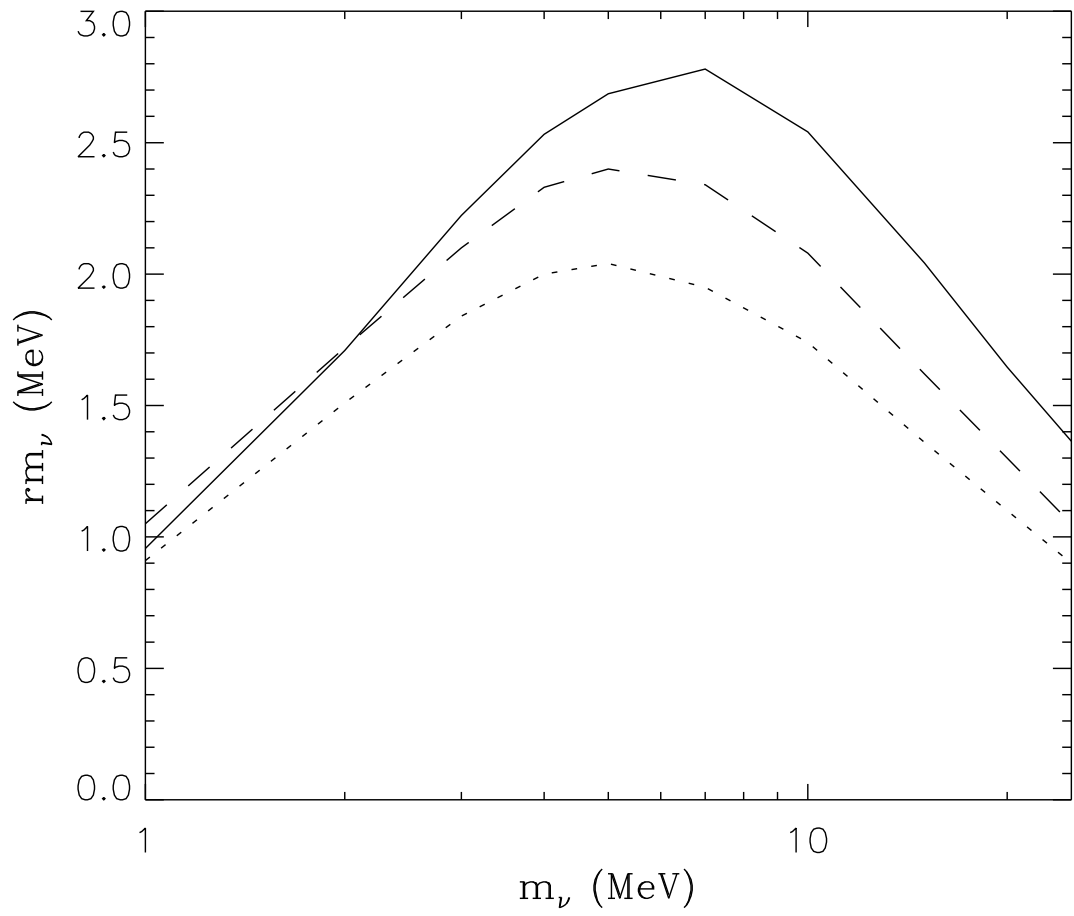
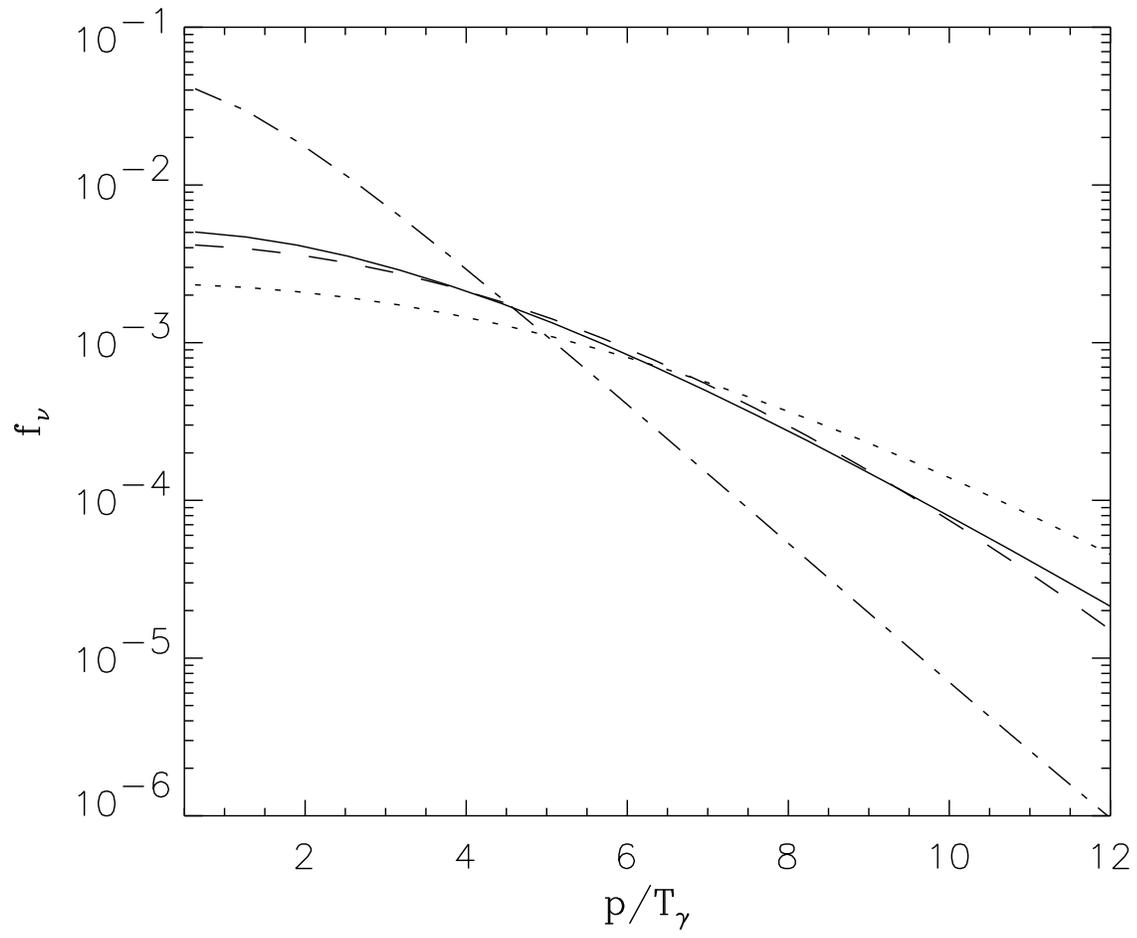
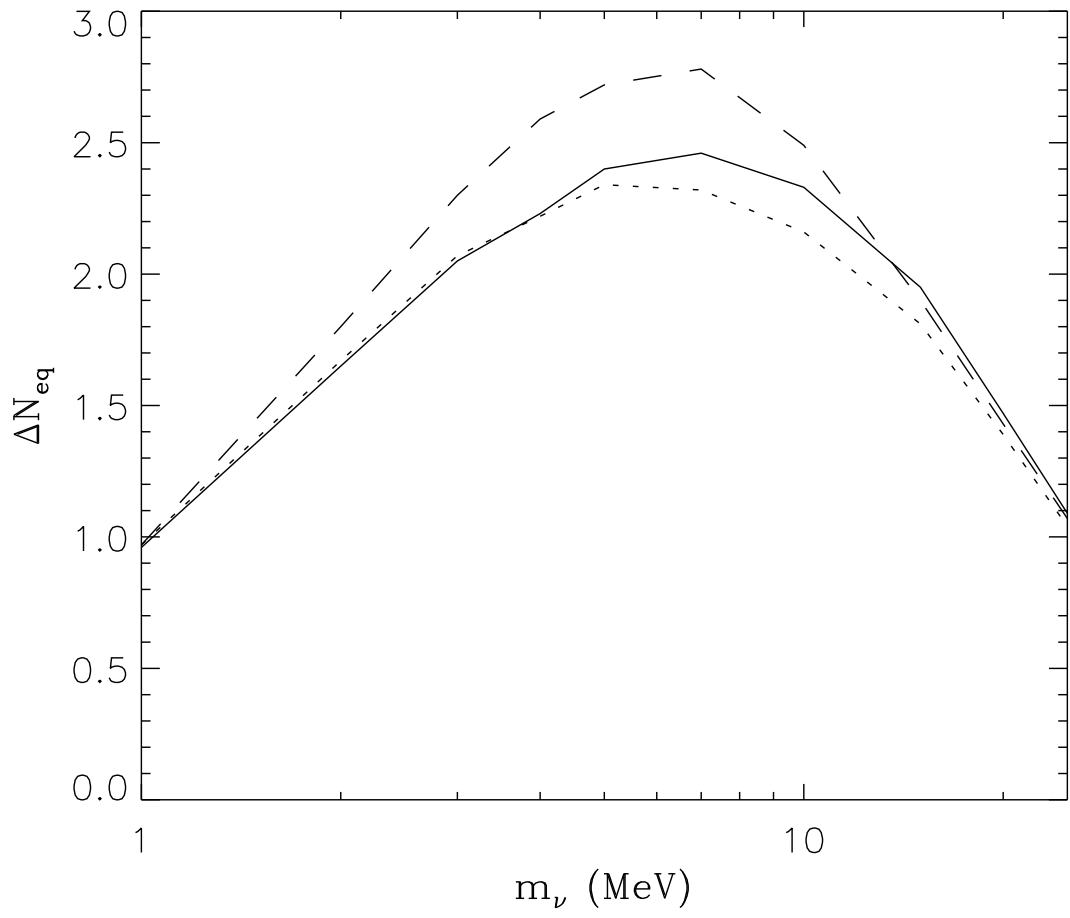
\end{document}